\documentclass[conference]{IEEEtran}
\IEEEoverridecommandlockouts

\usepackage{amsmath,amssymb,amsfonts}
\usepackage{algorithmic}
\usepackage{graphicx}
\usepackage{textcomp}
\usepackage{xcolor}
\def\BibTeX{{\rm B\kern-.05em{\sc i\kern-.025em b}\kern-.08em
    T\kern-.1667em\lower.7ex\hbox{E}\kern-.125emX}}

\definecolor{dg}{RGB}{64,64,64}

\usepackage[colorlinks = true,
		citecolor = black,
		urlcolor= blue,
		linkcolor=dg,]{hyperref}

\begin{document}

\title{Crypto-Economic Analysis of Web3 Funding Programs Using the Grant Maturity Framework
\thanks{This work is has been funded by the EU in the framework of the NGI TRUSTCHAIN project under the European Commission HORIZON-CL4-2022-HUMAN-01-03 grant: \href{https://cordis.europa.eu/project/id/101093274}{101093274} and \href{https://ens.domains/}{ENS Domains} through its \textit{Large Grants} program.}
}
\author{\IEEEauthorblockN{1\textsuperscript{st} Ben Biedermann}
\IEEEauthorblockA{\textit{Islands and Small States Institute} \\
\textit{University of Malta}\\
Msida, Malta \\
\href{https://orcid.org/0000-0003-1331-6517}{0000-0003-1331-6517}}

\and
\IEEEauthorblockN{2\textsuperscript{nd} Victoria Kozlova}
\IEEEauthorblockA{\textit{WID3\textsuperscript{+} Consortium} \\
\textit{ACURRAENT UG} \\
Frankfurt (Oder), Germany \\
\href{https://orcid.org/0000-0003-3303-3143}{0000-0003-3303-3143}}

\and
\IEEEauthorblockN{3\textsuperscript{rd} Fahima Gibrel}
\IEEEauthorblockA{\textit{Grant Innovation Lab} \\
\textit{Metagovernance Project Inc}\\
Brookline (MA), United States \\
\href{mailto:gibrel.fahima@gmail.com}{gibrel.fahima@gmail.com}}
}
\maketitle

\begin{abstract}
Web3 grant programs are evolving mechanisms aimed at supporting innovation within the blockchain ecosystem, yet little is known on about their effectiveness. This paper proposes the concept of \textit{maturity} to fill this gap and introduces the Grant Maturity Framework (GMF), a mixed-methods model for evaluating the maturity of Web3 grant programs. The GMF provides a systematic approach to assessing the structure, governance, and impact of Web3 grants, applied here to four prominent Ethereum layer-two (L2) grant programs: Arbitrum, Optimism, Mantle, and Taiko. By evaluating these programs using the GMF, the study categorises them into four maturity stages, ranging from experimental to advanced. The findings reveal that Arbitrum's Long-Term Incentive Pilot Program (LTIPP) and Optimism's Mission Rounds show higher maturity, while Mantle and Taiko are still in their early stages. The research concludes by discussing the user-centric development of a Web3 grant management platform aimed at improving the maturity and effectiveness of Web3 grant management processes based on the findings from the GMF. This work contributes to both practical and theoretical knowledge on Web3 grant program evaluation and tooling, providing a valuable resource for Web3 grant operators and stakeholders.
\end{abstract}

\begin{IEEEkeywords}
maturity model, Web3 governance, decentralised autonomous organisations (DAOs), crypto-economic systems, mixed-methods
\end{IEEEkeywords}

\section{Introduction}\label{sec_1}
Web3 grants are a relatively new phenomenon that has seen little standardisation. Thus, grant programs in Web3 lack a systematic framework for comparing and evaluating their outcomes. While some work has been done on Web3 grants in the context of creating a framework for decentralised science (DeSci)~\cite{ding_desci_2022} and grants found mention in research on Web3 governance~\cite{allen_exchange_2022}, no further reference to Web3 grants exists in the literature across all disciplines, including studies on procurement. This poses significant challenges to Web3 grantors, grant operators, grant-giving decentralised autonomous organisations (DAOs), and grantees because all actors rely solely on industry knowledge and informal guides.

In absence of a theoretical framework, the structure and benefits of decentralised grant programs cannot be reliably measured or tracked. This poses a problem for evaluating the utility of programs for allocating funding through distributed systems, specifically for using blockchain to disperse grants. Decentralised networks are said to be more open and less rigid than hierarchical and centralised structures~\cite{ding_desci_2022}, but their inaccessibility to outsiders creates information asymmetries both for applicants and funders.

Web3 grant programs and EU-backed grants address a diverse audience. The heavy use of jargon in Web3 grants, however, reduces applicant diversity and the impact of Web3 grants is skewed by applicant self-selection. Therefore, Web3 grant programs struggle to ascertain whether the population of applicants represents the desired range of fields or is a second order effect of the limitations in the design of the program. Even for long-standing grant initiatives, like those by the EU, measuring grant program effects is difficult and improved only recently~\cite{selebaj_effects_2021}. In the absence of formal, reliable, and complete data for Web3 grant programs, formal models for measuring capital efficiency cannot be applied~\cite{ odewole_capital_2020}.

To that end, this research proposes the grant maturity framework (GMF), which aims to fill the gap of a systematic framework that provides practitioners with ``a baseline and benchmark for [Web3 grant program] maturity and identifying areas for improvement''~ \cite{dener_govtech_2021}. The GMF builds on prior research by the World Bank Group’s (WBG) towards a government technology (GovTech) maturity index (GTMI). Thus, the GMF introduces an exploratory weighted composite framework, which is constructed using mixed methods for responding to three research questions.

\small
\begin{enumerate}
\item \textit{How can the structure and outcomes of Web3 grant programs be measured?}
\item \textit{What maturity levels do popular Web3 grant programs exhibit?}
\item \textit{Which lessons can be learned for the technical design of Web3 grant platforms?}
\end{enumerate}
\normalsize

The first research question aims at providing overview of the objectives, processes, and organisational structures that underpin Web3 grant programs. For responding to the second question, the maturity framework is applied to four concrete Web3 grant programs for assessing their maturity. Towards answering the third research question, this paper analyses the outcomes of implementing the results from the GMF within the scope of  AUTHBOND, an EU next generation internet (NGI) TRUSTCHAIN project that develops a grant funding platform according to the principles of a user-centric approach (UCA). This research problem is addressed by using mixed-methods for comparing the grant programs of four popular Ethereum layer-two (L2) ecosystems that have a token. As a result, the GMF serves as a toolkit for Web3 grant operators, program principals, and stakeholders for tracking the effects of Web3 grant programs and translating learnings into actionable insights for the development of grant funding platforms.

\section{Background}\label{sec_2}

This section provides an overview of the state of grant research in management science and Web3-specific literature. Grounded in the literature on public innovation funding and management science~\cite{albors_impact_2011,bartle_review_2003}, Web3 grants are understood as capital allocation using blockchain technology and funding its development. The emergence of novel grant mechanisms in Web3, such as \textit{quadratic funding}, illustrates the organisational shift from hierarchical structures to disintermediated crypto-economic systems Web3 stands for~\cite[p.~501]{shermin_disrupting_2017}. The cryptocurrency and Web3 sector is well known for heightened investment risks, thus, grants can take a key role in reducing the risk for capital allocators by providing funding incrementally and connecting it to non-financial support, such as technical assistance~\cite[p.~6]{gilbert_sustainable_2019}.

As a concept, grants fuse organisational concepts of Web3, such as decentralised autonomous organisations (DAOs) with conventional economic studies of the public and private sector~\cite{ding_desci_2022,monteiro_decentralised_2023,wang_self-sovereign_2020}. Following, DAOs are defined as “complex [...] smart contract [...] governance”~\cite[p.~501]{shermin_disrupting_2017} allowing for “human–machine hybrid [...] self-organization with no centralized hierarchy”~\cite[p.~1564]{ding_desci_2022}. In practice, the first DAO introduced the notion of “democratic investment fund[s]”~\cite[p.~4]{santos_dao_2018}, where DAOs combine established financial logic with disruptive technology and novel organisational structures. 

These organisations are now challenging the hegemonic structure of economic actors, most importantly the \textit{enterprise}~\cite{wang_novel_2024}. Stringently, DAOs may also take over the role of a procurement body or intermediary for innovation procurement in Web3. Innovation procurement intermediation is defined as “provid[ing] a link between at least two entities which need to connect in order to generate or adopt innovation”~\cite[p.~416]{edler_connecting_2016}. It is possible to transfer knowledge from conventional grant programs to Web3 grants by focusing on the communication with stakeholders. To this end, both conventional and Web3 grant programs create blog posts, hold webinars, publish social media posts, and disseminate educational materials. Thus, even Web3 grant programs cannot exclusively rely on blockchain rails, but at a minimum must communicate with the funding body and the grant recipients. This communicative role is fulfilled through conventional means rather than on-chain.

Grants are broadly defined as ``a financial donation awarded by the contracting authority to the grant beneficiary'', which can be tied to a specific action or unrestricted for objectives of a specific organisation~\cite{european_commission_grants_2023}. Operators and observers of government research and development (R\&D) grants have been analysing, evaluating, measuring, and criticising their impact~\cite{howell_financing_2017,lerner_government_2000}. It was found that in the case of the United States (US) Small Business Innovation Research (SIBR) government R\&D grants improved employment and monetisation capacities of recipients, however, did not lead to an increased likelihood of venture capital investments~\cite{lerner_government_2000}.

Thus, concerns over grant efficiency are not limited to Web3, but apply to many grant programs, both public and private. If funding efficiency does not distinguish Web3 grants from conventional grant programs, what makes Web3 grants different from public innovation procurement? While there is over forty years of evidence on the efficiency of the organisational structure for traditional innovation funding allocation~\cite[p.~4]{holmstrom_agency_1989}, organisational structures in Web3 are more volatile~\cite[p.~25]{zuo_development_2023}. Web3 only emerged around 2014, when the technology was available to ``embrace[...] a set of protocols based on blockchain, which intends to reinvent how to return data ownership to users and let everyone equally participate in it''~\cite[p.~4]{wan_web3_2023}.

\subsection{Web3 Grant Programs}\label{sec_2.1}

Although challenges for quantitatively measuring output and impact~\cite{ding_desci_2022,howell_financing_2017} are documented for grant programs in- and outside of Web3, the novelty of Web3 exacerbates these challenges. Based on the findings in~\cite{ding_desci_2022,wan_web3_2023} DAOs are an important organisational design in Web3 and a dominant feature of Web3 grant programs. The reliance of DAOs on smart contracts and decentralised technologies introduces additional challenges. This subsection specifies the context of the GMF as modelling the \textit{maturity} of governance technologies in the context of Web3 grant programs and draws from research by the World Bank Group (WBG).

Exploratory and inductive research on Web3 grants provides a broad overview of the Web3 grant landscape and common challenges, for example, outcome measurement, but does not follow a systematic approach~\cite{leventhal_state_2023_long,leventhal_state_2024}. The World Bank GovTech Maturity Index (GTMI) provides structure for mapping the relationship between resource governance and innovation output of Web3 grant programs. It is an example for how governance mechanisms can be associated with concrete technology outputs. The GMF adapts the GTMI framework for measuring the individual performance of participants in the quadrants ``core government systems, public service delivery, citizen engagement, and GovTech enablers''~\cite[p.~5]{dener_govtech_2021}.

When DAOs are used for grant delivery in Web3~\cite{austgen_dao_2023},  ``grants [...] directly surface the relationship between DAOs and traditional challenges in treasury management and public finance''~\cite[p.~27]{tan_open_2023}. Consequently, the concept of maturity allows the description of the evolution of grant-giving, from classical efficiency considerations~\cite{holmstrom_agency_1989} to organisational challenges posed by grants based on crypto-economic systems. Web3 grant programs typically encompass a blockchain network operator or backer, a program manager, applicants, and stake-holding communities~\cite{gilbert_sustainable_2019,howell_financing_2017}.

From an organisational perspective, Web3-specific programs differ from conventional grant programs in two aspects. Firstly, backers and program managers behind Web3 grant programs typically are positioned in the private sector and their organisational structure does not always conform to established governance structures, such as enterprises and public sector entities. Secondly, Web3 grant programs exist for a significantly shorter period compared to governmental grant programs. Blockchain and DAOs are ``governance technologies''~\cite{brekke_hacker-engineers_2021,shermin_disrupting_2017} , catalysing organisational experimentation and innovation in Web3 grant program design.
The openness and transparency of crypto-economic systems~\cite[p.~54]{santos_dao_2018} amplifies their effects. Web3 program operators can choose from an ever increasing set of governance mechanisms for the selection and distribution of grants, which is seen as a quality marker by Web3 participants~\cite{owocki_gitcoin_2024}. Yet, the growth and adoption of novel allocation mechanisms risks distracting from appropriately communicating and engaging with applicants and grantees.

\subsection{Grant Program Maturity}\label{sec_2.2}

Given the limitations of quantitative grant impact assessments, econometric measures, such as observing changes of transaction count, transaction velocity, and total value locked (TVL), are not enough. Therefore, the GMF measures Web3 grant program \textit{maturity} through a composite mixed-method framework. This approach adopts the WBG GTMI complementary positioning to traditional macroeconomic indicators for nation states, such as the gross domestic product (GDP), and monetarist metrics such as the quantity theory of money~\cite{sun_understanding_2004}.

Maturity models emerged from the intersection of innovation research and economics in the health care sector, evaluating the adoption of digital technologies~\cite{van_ede_assembling_2024,knosp_research_2018}. Grey literature and policy documents have applied and popularised the concept~\cite[see also]{dener_govtech_2021,queensland_audit_office_risk_2023}, but only \cite{kucinska-landwojtowicz_organizational_2023} provides an explicit definition of \textit{maturity}. Based on the state-of-the-art analysis by~\cite{kucinska-landwojtowicz_organizational_2023} and the application of maturity modelling for ``evaluat[ing] the current level of operational development''~\cite{yatskovskaya_integrated_2018}, Table~\ref{tab:grant_maturity} puts forth an explicit approach for defining maturity in the context of Web3 grant programs. Accordingly, maturity is understood ``in dynamic terms – as a process [...] [and] as a specific state or degree of perfection'' for measuring an organisation's success''~\cite[p.~62]{kucinska-landwojtowicz_organizational_2023}.

More stringently, for measuring Web3 grant programs structure and outcomes, maturity ``represents an anticipated, desired, or typical evolution path of these objects shaped as discrete stages''~\cite[p.~213]{becker_developing_2009}. The GMF operationalises this definition by classifying programs into four stages. Program properties are described through the definition of maturity stage, distinct program features, and relevant improvements for Web3 grant programs in the respective stage. It is implicit that any Web3 grant program must pass through one stage in order to reach the next one, which is grounded in the theory~\cite{yatskovskaya_integrated_2018}. Thus, the GMF incorporates a higher-better logic when scoring Web3 grant programs.

\begin{table}[htbp]
\caption{Stages of Maturity for Web3 Grant Programs}
\centering
\footnotesize
\begin{tabular}{p{2cm}p{6cm}}
\hline
\textbf{\textit{Maturity Stage}} & \textbf{\textit{Description}} \\
\hline
Experimental & Focus on exploring new funding mechanisms. Simple structure with limited stakeholder engagement. No formal processes or timelines. \\
\hline
Foundational & Basic governance and program structure defined. Initial mission, vision, and objectives. Simple application process and evaluation criteria. \\
\hline
Developmental & Improved structure with clearer application processes. Greater resource allocation and decision-making beyond core organisation. Focus on impact metrics and community feedback. \\
\hline
Advanced & Standardised processes with dedicated infrastructure and staff. Transparency, regular audits, and community engagement. Impact measurement tools and comprehensive decision-making. \\
\hline
\end{tabular}
\label{tab:grant_maturity}
\end{table}

Flexible and dynamic definitions of maturity have also led to criticism towards the concept~\cite[p.~8]{pereira_review_2020}. To address this shortcoming, the GMF links the conceptual maturity stages in Table~\ref{tab:grant_maturity} to a multivariate composite framework. The maturity model, thus, is adaptive through using organisational maturity~\cite{andersen_e-government_2006,johansson_roadmap_2019}, while providing replicable results based on the mixed-method operationalisation of an explicit and concise definition of the term.

\section{Methodology}\label{sec_3}

This research follows a mixed-method design to assess Web3 grant programs through the lens of their maturity, triangulate the findings, and suggest actionable improvements for Web3 grant tooling. The GMF combines the evaluation of the capability to undertake organisational improvements with the assessment of measures for increasing allocation efficiency by aggregating qualitative assessments and quantitative indicators. The findings were tested in the development of a novel Web3 grant platform, which included collecting qualitative feedback from industry. The methodology is structured in three phases. Firstly, expert assessments were collected using a questionnaire called \textit{rubric scoring framework}~\cite{biedermann_evaluating_2024}. Secondly, the rubrics and the nested indicators structured the collection of primary data from Web3 grant programs. Lastly, the GMF's maturity and rubric scores informed features of a Web3 grant platform, furthering the state-of-the-art in Web3 grant giving. This second round of qualitative data triangulated the findings of the GMF~\cite{creswell_designing_2017,datta_paradigm_2006}.

Primary data collection started with covering known challenges in conventional grant programs, such as capital efficiency, organisational design, and politico-economic stances of grant program operators~\cite{lerner_government_2000}. It was aimed to underpin the comprehensive definition of maturity by falsifiability~\cite[p.~17]{popper_objective_1973}. The selection of grant programs for this study was based on several criteria: the program must be active, involve both a DAO and a foundation, and focus on Ethereum Layer 2 solutions, excluding retrospective funding. Ultimately, the four grant programs Arbitrum, Mantle, Optimism, and Taiko were selected based on these criteria. Thus, the GMF verifies whether the concept of \textit{maturity} applies to Web3 grant programs~\cite[p.~7]{santos_dao_2018} and the qualitative statements from the expert rubric scoring hold true~\cite{hutton_abstraction_1990}.

In this regard, the study is based on the ``ontological assumption[s]''~\cite[pp.~94]{mukumbang_retroductive_2023} that Web3 grant programs exist to exercise ecosystem governance through the financial utility of the respective native token~\cite{beck_governance_2018,messias_understanding_2024}. As the GMF includes the financial assessment of budgets and grant amounts relative to the value of the native tokens for prices on the 2\textsuperscript{nd} September 2024 in United States Dollars on \cite{coinnarketcap_cryptocurrency_2024}, the GMF quantifies ``the value of communication'' as measure of maturity~\cite{johansson_roadmap_2019} in the final maturity score.

\subsection{Rubric Scoring Framework and Indicator Clustering}\label{sec_3.1}

Procurement of technology and software is documented in enterprises and the public sector~\cite{bartle_review_2003,johansson_roadmap_2019,uyarra_barriers_2014}, but only sporadically for DAOs~\cite{monteiro_decentralised_2023} and without taking the maturity of Web3 procurement systems and processes into account. Grant programs are scored holistically and based on a discrete set of categories. These scoring categories, were developed inductively and are the basis for the \textit{rubric scoring framework}, as well as the GMF. Broadly, the GMF consists of the six clusters defined in Table~\ref{tab:grant_rubric} below.

\begin{table}[htbp]
\caption{Web3 Grant Program Rubric Scoring Framework}
\centering
\footnotesize
\begin{tabular}{p{3.8cm}p{4.5cm}}
\hline
\textbf{\textit{Rubric}} & \textbf{\textit{Assessment Criteria}} \\
\hline
Focus Areas, Objectives (FAO) & Defined focus areas and strategic goals \\
\hline
Program Structure (PSO) & Clarity of decision-making processes \\
\hline
Governance (GOV) & Governance structure, transparency, funding sources \\
\hline
Effectiveness, Impact (EFI) & Explicit goals, success criteria, impact assessment \\
\hline
Transparency (TAC) & Transparent, accountable reporting mechanisms \\
\hline
Community Engagement (COM) & Community reach, engagement, incentives \\
\hline
\end{tabular}
\label{tab:grant_rubric}
\end{table}

According to the rubric scoring framework a questionnaire was designed and shared among five participants, who were representatives of zkSync, Arbitrum, Octant, Balancer, and GitCoin, as well as Web3 grant professionals who are not currently affiliated with a grant-giving entity in Web3. Using the Delphi method, as suggested for health management science~\cite{van_ede_assembling_2024}, the appraisal scheme required evaluators to score programs in the categories and subcategories on a continuous scale from one (1) to five (5). A low score represents low maturity, a score of three (3) signals intermediate maturity, and five (5) signifies high perceived maturity. Evaluators were requested to add comments for each score given. The scores of the subcategories were then aggregated to category scores and a written justification was given for the overall evaluation. The qualitative results instructed the researchers in the selection of available quantitative data points to be included as indicators in the GMF for scoring the selected grant programs. Quantitative scores were included as indicators for each rubric in the GMF.

\subsection{Construction of the Grant Maturity Framework}\label{sec_3.2}

The indicator composition was informed by the results of Delphi study, consisting of 46 data points that are grouped into six rubrics clusters (Table~\ref{tab:grant_rubric}). A subset of 40 data points are included in the index with a non-zero weight, the remaining six (6) were taken into account to provide a ``rich description''~\cite[p.~197]{mcbride_sailing_2019} of selected Web3 grant programs in the dataset, but are not reflected in the GMF score. All non-zero weighted indicators are normalised and aggregated using equal weights.

As a result of the refinement process, the composite index consists of two main components. Namely, the rubric maturity scores for all analysed Web3 grant programs and the GMF as the normalised composite of all rubric maturity scores. For calculating the GMF maturity score, first, the data was collected and listed according to the structure in Table~\ref{tab:grant_rubric}. The data was normalised through the min-max normalisation function to ensure comparability across variables and construct validity of the composites, i.e. “[a]void adding up apples and oranges”~\cite[p.~27]{oecd_handbook_2008}. Min-max normalisation was selected over other normalisation methods because it is highlighted in literature that describes the governance of uncertainty~\cite{winters_when_2004}. Third, the rubric maturity scores were calculated, normalised, and aggregated to form the final GMF composite. Based on this calculation, the maturity stages in Table~\ref{tab:grant_maturity} were defined as quartiles from zero to one.\\

\subsubsection{Composite Rubric Score Calculation}\label{sec_3.2.1}

The composite rubric score for indicators of one rubric for each program was calculated using the formula
\footnotesize
\[
CRS_{ik} = \sum_{j=1}^{n} w_{jk} X'_{ijk}
\]

where:

\begin{itemize}
    \item \( CRS_{ik} \) is the composite indicator for the \( k \)-th rubric score of the \( i \)-th program,
    \item \( X'_{ijk} \) is the normalised value of the \( j \)-th variable for the \( k \)-th rubric score of the \( i \)-th program,
    \item \( w_{jk} \) is the weight assigned to the \( j \)-th variable for the \( k \)-th rubric score of the \( i \)-th program, implying equal weights, \( w_{jk} = \frac{1}{n} \),
    \item \( n \) is the number of variables for a given rubric score of a given program.
\end{itemize}\vspace{7pt}
\normalsize
\subsubsection{Composite Grant Maturity Framework Score}\label{sec_3.2.2}
The composite GMF score for rubric scores was calculated using the formula
\footnotesize
\[
GMF_i = \sum_{k=1}^{m} w_k CRS'_{ik}
\]

where:

\begin{itemize}
    \item \( GMF_i \) is the composite indicator for the \( i \)-th program,
    \item \( CRS'_{ik} \) is the normalised value for the \( k \)-th rubric score of the \( i \)-th program,
    \item \( w_k \) is the weight assigned to the \( k \)-th rubric score of the \( i \)-th program, implying equal weights, \( w_k = \frac{1}{m} \),
    \item \( m \) is the number of rubric scores for a given program.
\end{itemize}\vspace{7pt}
\normalsize

\subsubsection{Construct Validity and Reliability}\label{sec_3.2.3}

Indicators were grouped and nested into rubrics, decoupling and abstracting the overall framework from the underlying indicators. Public data was used to allow for the replicability of the construction of the framework. Yet, the index currently assumes that all indicators and rubric maturity scores are equally important because of equal weights. Further research may relativise this stance and introduce rubric weights with \( 0 \leq w_{ij} \leq 1 \) and \( \sum_{i,j=1}^{m,n} w_{ij} = 1\).

\section{Results}\label{sec_4}

This section presents the results of the analysis conducted on popular Web3 grant programs using the GMF. The focus is on examining the maturity levels of the set of Web3 grant programs, consisting of Arbitrum (ARB), Optimism, Mantle, and Taiko, as well as evaluating their structures, processes, and outcomes. The results also aim to answer the research questions posed in section~\ref{sec_1} by reviewing the rubric scores and maturity levels in~\ref{sec_4.1}, programs in~\ref{sec_4.2}, and describing feedback from improved Web3 grant platform in~\ref{sec_4.3}.

\subsection{Measuring Web3 Grant Program Maturity}\label{sec_4.1}

The structure and outcomes of Web3 grant programs can be measured using 40 indicators across the six rubric categories in Table~\ref{tab:grant_rubric}. Each rubric category includes multiple indicators that evaluate aspects of the program, such as grant size, evaluation criteria, organisational structure, governance mechanisms, and community involvement. The indicators in the FAO rubric assess grant types, funding methods, and evaluation timeframes, among other factors. The PSO rubric evaluates the organisational structure of the grantor and the allocation of funds. The GOV rubric includes data points on the program's mission and vision documents and its alignment with objectives like network growth or philanthropy. The EFI rubric measures how effectively the program achieves its stated goals, while the TAC rubric focuses on transparency and accountability in decision-making processes. Finally, the COM rubric looks at the level of community engagement, including the applicant count, average grant size, and program management team size. In aggregate, the rubrics form the overall Web3 grant maturity scale displayed in Table~\ref{tab:revised_maturity_stages}.

\begin{table}[htbp]
\caption{Refined Maturity Stages}
\centering
\footnotesize
\begin{tabular}{p{1.5cm}p{2.5cm}p{3.5cm}}
\hline
\textbf{\textit{Maturity}} & \textbf{\textit{Quartile Range}} & \textbf{\textit{Description}} \\
\hline
Experimental & \( 0 \leq GMF < 0.25 \) & Exploratory with limited structure and governance \\
Foundational & \( 0.25 \leq GMF < 0.5 \) & Programs start to define objectives, structures, and governances \\
Developmental & \( 0.5 \leq GMF < 0.75 \) & Clear structures are in place with defined processes for allocation \\
Advanced & \( 0.75 \leq GMF \leq 1.0 \) & Programs have robust and transparent governance with impact measurement \\
\hline
\end{tabular}
\label{tab:revised_maturity_stages}
\end{table}

\subsection{Maturity Levels of Popular Web3 Grant Programs}\label{sec_4.2}

Based on the selection of the four most relevant Web3 grant programs, the GMF contains six observations because Arbitrum as one of the largest grant programs already differentiated between multiple tracks. These three ARB sub-programs were in place concurrently and too different in scope and nature to be included as one. Thus, the final set of assessed Web3 grant programs is the Arbitrum Short Term Incentive Program (STIP)~\cite{tnorm_arbitrums_2023,arbitrum_dao_arbitrum_2023} including its capital increase called \textit{Backfund}~\cite{frisson_arbitrum_2023}, the continuation of Arbitrum STIP referred to as \textit{Bridge}~\cite{lumley_stip-bridge_2024,arbitrum_dao_double-down_2024}, the Arbitrum Long-Term Incentive Pilot Program (LTIPP)~\cite{stein_arbitrum_2024}, the Mantle Grants Program~\cite{bitdao_passed_2023}, the Optimism Mission Round Program including rounds 3 and 4~\cite{the_optimism_collective_what_2024}, and Taiko's Incentivisation Grant Program. The maturity levels of these programs are detailed in Table \ref{tab:gmf_composite_score}, which provides both additive and normalised composite scores.

\begin{table}[htbp]
\caption{GMF Composite Scores}
\centering
\footnotesize
\begin{tabular}{p{2.7cm}p{2cm}p{2.3cm}}
\hline
\textbf{\textit{Grant Program}} & \textbf{\textit{Additive} \( GMF\)} & \textbf{\textit{Normalised} \( GMF \)} \\
\hline
ARB STIP \& Backfund & 3.6653 & 0.4349 \\
ARB STIP Bridge & 3.7582 & 0.5251 \\
ARB LTIPP & 4.2679 & 0.6755 \\
Mantle & 3.1787 & 0.2729 \\
Taiko & 3.1191 & 0.2334 \\
Optimism & 3.6397 & 0.6105 \\
\hline
\end{tabular}
\label{tab:gmf_composite_score}
\end{table}

The programs were chosen as they represent considerable capital- and market share across Web3 ecosystems, even more so for Ethereum specifically~\cite{marz_airdrops_2024} and cover a variety of maturity levels, from experimental to advanced stages. In the study, ARB LTIPP program achieved the highest score with a normalised composite in relative notation of 67.55\%, placing it in the \textit{developmental} stage corresponding to the third quartile. The Optimism Mission Rounds, with a normalised composite score of 61.05\%, also fall within the \textit{developmental} stage but with slightly lower maturity than ARB LTIPP. In contrast, the Taiko and Mantle programs scored lower, with normalised composites of 23.34\% and 27.29\% respectively, which places them in the \textit{experimental} stage, indicative of their early-phase status and need for further development. The detailed rubric maturity scores for each program are presented in Table~\ref{tab:rubric_scores}.

\begin{table}[htbp]
\caption{Normalised Rubric Scores of Analysed Programs}
\begin{center}
\footnotesize
\begin{tabular}{p{2.9cm}p{0.5cm}p{0.5cm}p{0.5cm}p{0.5cm}p{0.5cm}p{0.5cm}p{0.5cm}}
\hline
\textbf{\textit{Grant Program}} & \textbf{\textit{FAO}} & \textbf{\textit{PSO}} & \textbf{\textit{GOV}} & \textbf{EFI} & \textbf{\textit{TAC}} & \textbf{\textit{COM}} \\
\hline
ARB STIP \& Backfund & 0.6856 & 1.0000 & 1.0000 & 0.0769 & 0.5455 & 0.2207 \\
ARB STIP Bridge & 1.0000 & 0.9184 & 0.9932 & 0.0000 & 0.5455 & 0.0000 \\
ARB LTIPP & 0.1987 & 0.9892 & 0.9932 & 0.6923 & 0.9091 & 1.0000 \\
Mantle & 0.1477 & 0.6494 & 0.4694 & 0.6154 & 0.0000 & 0.3834 \\
Taiko & 0.0000 & 0.7347 & 0.0000 & 0.0000 & 0.0000 & 0.2040 \\
Optimism & 0.2210 & 1.0000 & 0.0000 & 1.0000 & 0.0000 & 0.4419 \\
\hline
\end{tabular}
\end{center}
\label{tab:rubric_scores}
\end{table}

This table shows the normalised scores for each program across the six rubric categories. Each category is scored on a scale from 0 to 1, which can also be expressed in relative notation through percentages, with higher scores indicating better performance in that particular aspect of the program. Program-specific results are presented in the following.

\subsubsection{Optimism Growth and Experiment Grants}\label{sec_4.2.1}
The Optimism Mission Grants program, part of the larger Optimism grant initiative, scored a normalised composite of 61.05\%, placing it in the \textit{developmental} stage. The program’s objective is to drive network adoption and support applications that align with the goals of the Optimism Collective. This program has matured beyond its initial exploratory phase, with well-established evaluation criteria and some degree of transparency in its decision-making process. Yet, there is still potential to expand its impact by diversifying supported projects and improving its governance framework.

\subsubsection{Arbitrum Grant Programs}\label{sec_4.2.2}
Arbitrum’s multiple grant programs, including the STIP, STIP Backfund, STIP Bridge, and LTIPP, represent a range of maturity levels. The STIP and STIP Backfund programs scored normalised composites of 43.49\% and 52.51\%, respectively, indicating their position in the \textit{foundational} stage. These programs are designed to provide short-term incentives for network growth, but they lack some of the more advanced structures and processes seen in the LTIPP program. The Arbitrum LTIPP program, with a normalised composite of 67.55\%, is in the \textit{developmental} stage, indicating its more advanced maturity. The LTIPP program focuses on long-term incentives and has a more developed governance structure, well-defined objectives, and measurable impacts.

\subsubsection{Taiko Labs Grants}\label{sec_4.2.3}
The Taiko Labs Incentivisation Grant Program scored a normalised composite of 23.34\%, placing it in the \textit{experimental} stage. While the program has ambitious goals for scalability and network adoption, it is still in the early stages of implementation. The program has made progress in defining its objectives but is still lacking in some areas such as governance, transparency, and community engagement.

\subsubsection{Mantle Grants}\label{sec_4.2.4}
The Mantle Grants Program, scored a normalised composite of 27.29\%, placing it in the \textit{foundational} stage. While the program has a clear structure, it is still developing its governance processes and transparency mechanisms. To move into the \textit{developmental} stage, the program will need to refine its evaluation processes, improve the visibility of its funding decisions, and enhance its engagement with the broader community.

\subsection{Technology-Driven Improvement of Web3 Grant Maturity}\label{sec_4.3}

The iterative development of the platform, guided by user feedback, led to improvements in grant management workflows when compared to the tooling used by programs assessed by the GMF, such as milestone tracking, and proposal submission. Given the large variance in results for the EFI rubric score, validation sessions, including user interviews and stakeholder brainstorming, particular attention was paid when participants identified inefficiencies in fund disbursement in the grants landscape. This prompted the introduction of a milestone-based funding system where grant issuers can review submitted milestones and release payments accordingly.

\begin{figure}[htbp]
\centerline{\includegraphics[scale=0.24]{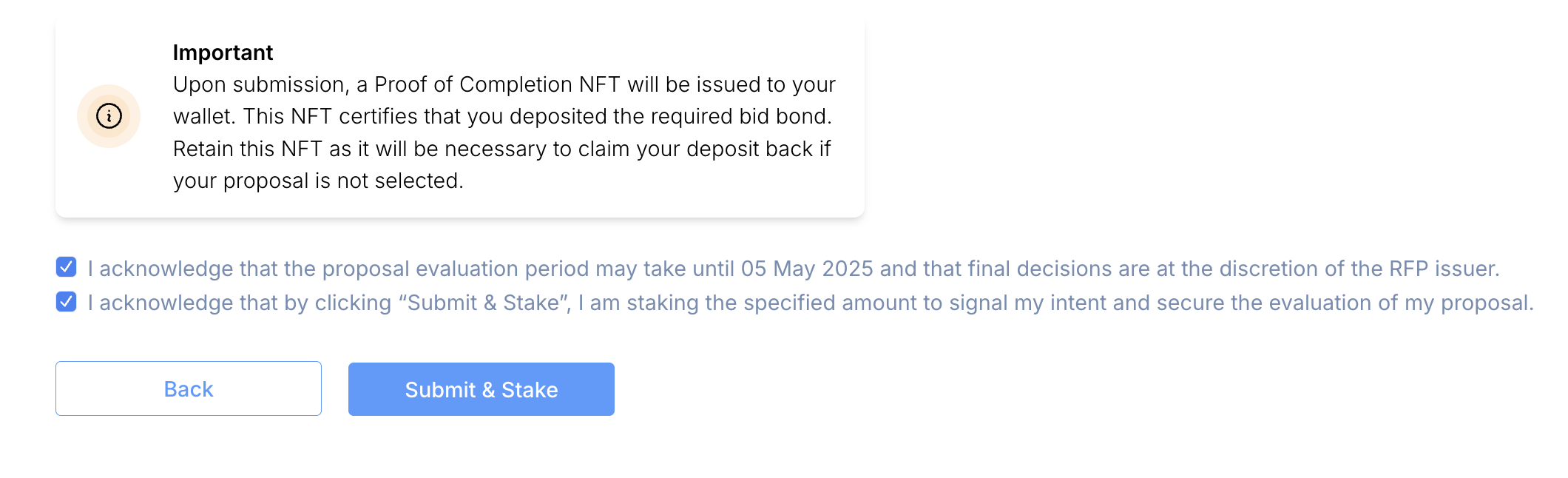}}
\caption{One step staking and submission flow.}
\label{fig:stake-submit}
\end{figure}

\begin{figure}[htbp]
\centerline{\includegraphics[scale=0.4]{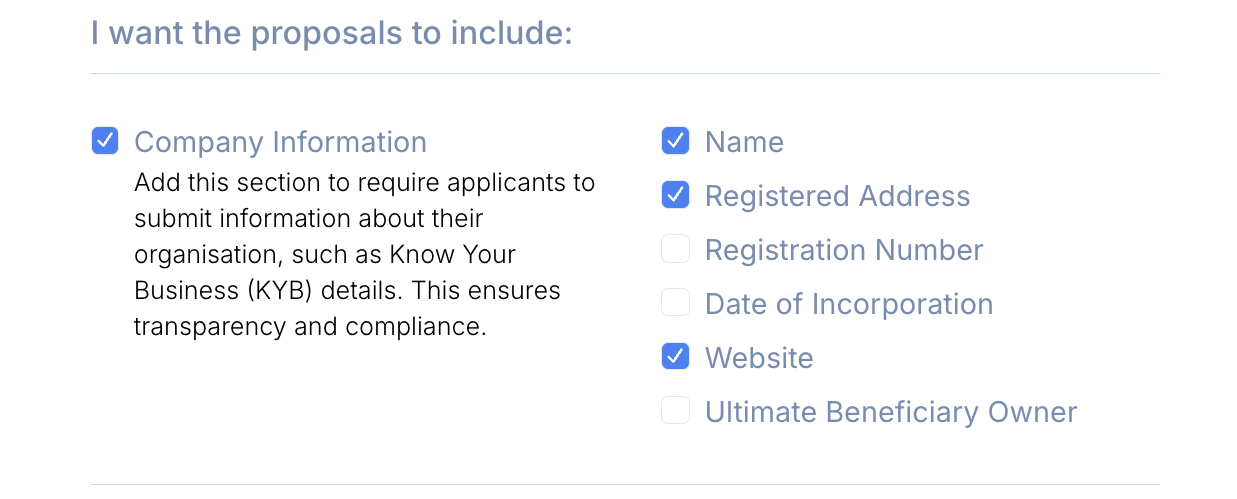}}
\caption{Optional KYB form in application flow.}
\label{fig:kyb-optional}
\end{figure}

Following, overall low scores in TAC, a streamlined proposal submission process was developed for the prototype of a Web3 grant platform, combining bid bond staking and proposal submission into a single blockchain transaction. Figure~\ref{fig:stake-submit} shows how the prototype was designed to reduce complexity for applicants. Additionally, aspects not covered in the GMF were tested as an optional feature based on user feedback, such as know your business (KYB) verification in Figure~\ref{fig:kyb-optional}. The same applies to bid bonds, allowing Web3 grant operators to disable this requirement if deemed unnecessary.

User profiles and dashboards were implemented to enhance transparency, usability, and create a place for community engagement. The absence of user profiles in earlier iterations led to concerns regarding information asymmetry. Finally, on-platform evaluation tools were also introduced to improve grant assessment capabilities, so that funders manage milestone reviews, provide structured feedback, and conduct qualitative and quantitative evaluations.

\section{Discussion}\label{sec_5}

The results align with the positioning of the grant programs in the Web3 market. For example, Optimism emphasises collective decision-making but its low governance score highlights the risks of decentralised governance, where outsiders may be left behind. Similarly, Arbitrum’s stratification of sub-programs reflects partisan DAO voting, excluding the wider community. Taiko’s score aligns with its position as a newer grant program, while Mantle is affected by intransparent governance and personnel overlap, justifying their lower scores.

The findings underscore the GMF’s ability to bridge exploratory research and substantial analysis~\cite{mukumbang_retroductive_2023}, similar to studies on executive compensation~\cite{billett_stockholder_2010}, network value~\cite{papaioannou_business_2023}, and trust models for voting~\cite{baranski_trust-centric_2024,lin_voting_2023}. The GMF highlights the importance of organisational maturity and transparency, enhancing legitimacy among expert audiences~\cite[p.~116]{curtin_does_2006} and rapport in Web3 grant processes~\cite{suddaby_legitimacy_2017}.

In sum, the GMF serves as a micro-level maturity index for Web3 grants. The GMF is a non-exhaustive collection of indicators, providing a foundational model for assessing the maturity of selected Web3 grant programs. Further research should refine the tracked characteristics for a larger set of programs, including retroactive ones and those outside the Ethereum network. This research follows OECD guidelines on multivariate composite indicators~\cite{oecd_handbook_2008}, but lacks the maturity of long-standing macro-economic composite indices, such as the human development index (HDI).

\section{Conclusions and Future Work}\label{sec_6}

The GMF contributed not only on a practical level to the understanding of Web3 grant programs, but also laid the foundation for future theoretical work that could challenge the legitimacy of processes in incumbent Web3 grant programs based on their maturity. The study was designed to ascertain the characteristics, differences, and applications of Web3 specific grant programs. In turn, the GMF may accelerate the maturing of Web3 grant programs by providing Web3 grant operators a tool for self-assessment and benchmarking. As a framework, it already informed the approach of a concrete development initiative for Web3 grant tooling. As a result, the research could respond to the three research questions by defining the GMF, applying it to four Web3 grant programs, and technically addressing shortcomings that led to lower maturity scores. In the future, the GMF may be used as a tool for Web3 grant program strategy development and audits, thereby increasing the robustness and integrity of Web3 grants.

\section*{Acknowledgment}

We thank Matthew Scerri for his contributions to the AUTHBOND platform, Eugene Leventhal for his feedback, the anonymous reviewers, and the participants in the study for their support.

\bibliographystyle{IEEEtran}
\bibliography{IEEEabrv,ieee_icbc_tc_bib_cleaned_proper}

\end{document}